\documentstyle[12pt]{article}
\date{}
\begin{document}
\title{A Phenomenological Expression for Deuteron Electromagnetic
Form Factors Based on Perturbative QCD Predictions\thanks{The Project
Supported by the National Science Foundation of China (NSFC)and Grant of
Academia Sinica.}}
\author{}
\maketitle

\begin{center}
Hui-fang WU$^{1),2)}$,Jun CAO$^{1)}$ and Tao HUANG$^{1)}$ \\
~~~~~1) Institute of High Energy Physics, P.O. Box 918(4),
Beijing, 100039, China.\\
and \\
2) Institute of Theoretical Physics,P.O.Box 2735,
Beijing, 100080, China.  
\end{center}

\begin{abstract}
For deuteron electromagnetic form factors,perturbative QCD(pQCD) predicts
that $G^{+}_{00}$ becomes the dominate helicity amplitude and that 
$G^{+}_{+0}$ and $G^{+}_{+-}$ are suppressed by factors $\Lambda_{\rm QCD}/Q$
and $\Lambda_{\rm QCD}^2/Q^2$ at large $Q^2$,respectively. We try to discuss
the higher order corrections beyond the pQCD asymptotic predictions by
interpolating an analytical form to the intermediate energy region.
From fitting the data,our results show that the helicity-zero to
zero matrix element $G^{+}_{00}$ dominates the gross structure function
$A(Q^2)$ in both of the large and intermediate energy regions;  it is a good
approximation for $G^{+}_{+-}$ to ignore the higher order contributions and
the higher order corrections to $G^{+}_{+0}$ should be taken into account
due to sizeable contributions in the intermediate energy region.
\end{abstract}



\newpage
\section{Introduction}

{\hskip 0.6cm}It was found a long time ago that the traditional meson-nucleon picture can not
explain the form factors of the deuteron as the momentum transfer $Q^2>1$ 
GeV$^{2~[1]}$. It means that the fundamental degrees of freedom 
of QCD, the quark and gluon degrees of freedom, must be taken into account.
However, A pure perturbative QCD (PQCD) calculation$^{[2]}$  shows that
the theoretical prediction is  much smaller than the data at currently
accessible energies, although it may be correct in very large $Q^2$. 
To explain the deuteron form factors in the intermediate energy region, 
we have suggested a QCD-inspired model for the helicity-zero
to zero matrix elements $G^{+}_{00}$ in the light-cone frame$^{[3]}$,
which should be the dominant amplitude from PQCD predictions$^{[4]}$.
This model can explain the data of the deuteron electromagnetic
structure function  $A(Q^2)$ and shows that $G^{+}_{00}$ is already dominant at
$Q^2$ of 1 GeV$^2$. Furthermore, it was found that $G^{+}_{+0}$ can not be
neglected in the form factor $G_M(Q^2)^{[5,6]}$ and $G^{+}_{+-}$ plays an
important role in $G_Q(Q^2)^{[6]}$. Neglecting them will result in
contradictions with both the data and the conventional meson-nucleon picture
in the low energy region.

\vspace{0.5cm}

Perturbative QCD (PQCD) predicts that $G^{+}_{00}$ becomes the dominant helicity
amplitude at large $Q^2$ and that $G^{+}_{+0}$ and $G^{+}_{+-}$ are suppressed by
factors $\Lambda_{\rm QCD}/Q$ and $\Lambda_{\rm QCD}^2/Q^2$, respectively.
neglecting $G^{+}_{+0}$ and $G^{+}_{+-}$ contributions at large $Q^2$, we have the
relation approximately,
$$
G_C:G_M:G_Q=(1-\frac{2}{3}\eta):2:-1 , \eqno (1)
$$
where $\eta=Q^2/4M^2$ and $M$ is the mass of the deuteron. However, the 
helicity-flip amplitude $G^{+}_{+0}$ and $G^{+}_{+-}$ contribute to $G_M$ and $G_Q$
because of the kinematic enhancement in the intermediate energy region.
In order to
explore the role of $G^{+}_{+0}$ and $G^{+}_{+0}$, we have tried to expand them 
to the second order in $\Lambda_{\rm QCD}/M^{[6]}$, according to the
QCD predictions at large $Q^2$. This expansion can connects smoothly 
PQCD predictions in the high energy region with traditional nuclear
physics predictions in the low energy region. 
It was shown that the second order
contribution strongly affects the behavior of $G_Q$ in the intermediate
energy region. At large $Q^2$, the ratio of form factors (1)  
is slightly modified.

\vspace{0.5cm}

Following this approach, $G^{+}_{+0}$ and $G^{+}_{+-}$ will be expanded to higher orders
(beyond the second order) in $\Lambda_{\rm QCD}/M$ according to the PQCD
prediction at large $Q^2$. In order to explore the role of higher order
contributions we discuss the possibility to interpolate an expression
for $G^{+}_{+0}$ and $G^{+}_{+-}$ to the intermediate energy region in
this paper. It is worthwhile to
unify the predictions for the deuteron form factors from the low energy to
large energy region.

\vspace{0.5cm}

A general consideration based on the PQCD predictions for the deuteron from
factors is given in Sec. 2. As an example, a phenomenological analytic form
including the higher order corrections in $\Lambda_{\rm QCD}/M$ is suggested
in Sec. 3. The numerical results and summary are presented in Sec. 4
and Sec. 5, respectively.

\section{A General Consideration Based on the PQCD Predictions}

{\hskip 0.6cm}For the deuteron case, the matrix element of the electromagnetic current
$J^\mu$ is defined as
$$
G_{\lambda^\prime\lambda}^\mu=\langle
P^\prime \lambda^\prime \mid J^\mu \mid P\lambda\rangle ~,  \eqno (2)
$$
where $Q^2=-(P^\prime-P)^2$ and $|P\lambda\rangle$ is an eigenstate of the
deuteron
with momentum $P$ and helicity $\lambda$. In the standard light-cone frame
(LCF), defined by Ref.[7] $q^+=0, q_y=0$, and $q_x=Q$,
the charge, magnetic, and quadrapole form factors can be obtained from 
the plus component of three
helicity matrix elements:
$$
{\hskip 0.7cm}G_C  = \frac{1}{2p^+(2\eta+1)}\left[ (1-\frac{2}{3}\eta) G_{00}^+ +
\frac{8}{3}\sqrt{2\eta}  G_{+0}^+ +\frac{2}{3}(2\eta-1) G_{+-}^+ \right] ,
\eqno (3a)
$$

$$
G_M  =   \frac{1}{2p^+(2\eta+1)}\left[ 2 G_{00}^+ +\frac{2(2\eta-1)}
{\sqrt{2\eta}}  G_{+0}^{+} -2 G_{+-}^+ \right] ~~~~~~~~~~~~~~~~ \eqno (3b)
$$

and

$$
G_Q  = \frac{1}{2p^+(2\eta+1)}\left[ - G_{00}^+ +\sqrt{\frac{2}{\eta}}
 G_{+0}^{+} -\frac{\eta+1}{\eta} G_{+-}^+ \right] . ~~~~~~~~~~~~~~~~ \eqno (3c) 
$$
In terms of $G_C$, $G_M$, and $G_Q$, the Rosenbluth cross section 
and the tensor polarization $T_{20}$ for elastic $ed$ scattering can be
expressed as
$$
\frac{d\sigma}{d\Omega}=\left( \frac{d\sigma}{d\Omega} \right)_{\rm Mott}
\left[ A(Q^2)+B(Q^2)\tan^2(\frac{\theta}{2}) \right]  \ , ~~~~~~~~\eqno (4)
$$
and
$$
{\hskip 1.2cm}T_{20}=-\frac{
\frac{8}{9}\eta^2G_Q^2+\frac{8}{3}\eta G_C G_Q+\frac{2}{3}\eta 
G_M^2 \left[\frac{1}{2} +(1+\eta)\tan^2(\frac{\theta}{2})\right]
}{\sqrt{2}\left[A+B\tan^2(\frac{\theta}{2})\right]} , \eqno (5)
$$
where $A(Q^2)$ and $B(Q^2)$ are given by

$$
A(Q^2)  =  G_C^2+\frac{2}{3} \eta G_M^2+\frac{8}{9}\eta^2 G_Q^2  \eqno (6)
$$
and
$$
B(Q^2)  =  \frac{4}{3}\eta (1+\eta )G_M^2  . ~~~~~~~~~~~~\eqno (7)
$$

\vspace{0.5cm}

It was shown $^{[4]}$ that, in LCF, the helicity-zero to zero matrix element
$G^{+}_{00}$ would be the dominant helicity amplitude at large $Q^2$ for elastic
$ed$ scattering from the PQCD predictions. It means the $G^{+}_{00}$ dominance in
the structure function $A(Q^2)$. It was also argued$^{[8]}$ 
that the dominance of $G^{+}_{00}$ begins at $Q^2\gg 2M \Lambda_{\rm QCD}\sim 0.8$ 
GeV$^2$ but not $\eta>>1$. Thus
$2M\Lambda_{\rm QCD}$ is a scale of validity of PQCD predictions and the
quark and gluon degrees of freedom in the deuteron should be taken into
account to solve the problem that the experimental results of $A(Q^2)$ are
in sharp disagreement with the meson exchange calculations for $Q^2>0.8$
GeV$^{2 [1]}$. To make detailed prediction for $G^{+}_{00}$, we have
suggested a QCD-inspired model$^{[3]}$ in the region of $Q^2>1$ GeV$^2$
based on the reduced form factor method$^{[9]}$, which fit the data well.

\vspace{0.5cm}

PQCD predicts that $G^{+}_{+0}$ and $G^{+}_{+-}$ are suppressed by
factors $\Lambda_{\rm QCD}/Q$ and $\Lambda_{\rm QCD}^2/Q^2$, respectively.
However, in the intermediate energy region, $G^{+}_{00}$ dominates the charge
form factor $G_C$, but not $G_M$ and $G_Q$. As shown in Eq.(3), 
while $\eta<\frac{1}{2}$, the $G^{+}_{+0}$ contribution to $G_M$ and
$G_Q$ are enhanced by a factor $\frac{1}{\sqrt{2\eta}}$ and $G^{+}_{+-}$ 
contribution to $G_Q$ is enhanced by a factor $\frac{1}{2\eta}$.
Although $G^{+}_{+0}$ and $G^{+}_{+-}$ are suppressed for dynamic reason, they 
contribute significantly to $G_M$ and $G_Q$ because of the kinematic
enhancement. Without these contributions, the predicted form factors, 
except for $G_C$, are in sharp disagreement with the data. 
In order to explore the role of $G^{+}_{+0}$ and $G^{+}_{+-}$,
we interpolate a general expression based on perturbative QCD predictions,
$$
G^{+}_{+0} = \frac{1}{\sqrt{2\eta}} g_{+0}(\eta) ~G^{+}_{00} ~~
\eqno (8a) 
$$

$$
G^{+}_{+-} = \frac{1}{2 \eta} g_{+-}(\eta) ~G^{+}_{00}~~,
\eqno (8b) 
$$
where $g_{+0}(\eta)$ and $g_{+-}(\eta)$ are any functions of $\eta$ with
$\eta \equiv \frac{Q^{2}}{4M^{2}}$. Obviously $G^{+}_{+0}$ and $G^{+}_{+-}$
are suppressed by factors $\Lambda_{QCD}/Q$ and $\Lambda^{2}_{QCD}/Q^{2}$ as
long as $g_{+0}(\eta)$ and $g_{+-}(\eta)$ satisfy the following condition,

$$
g_{+0}(\eta), ~~g_{+-}(\eta) \rightarrow O(1)~~~~ as ~~~~\eta
\rightarrow \infty ~~.  \eqno (9)
$$
Substituting Eqs.(8) into Eqs.(3), one can get

$$
G_{c} = \frac{1}{2p^{+}(2\eta + 1)} [(1 - \frac{2}{3} \eta) + \frac{8}{3} ~
g_{+0}(\eta) + \frac{2}{3} (2\eta - 1) \frac{1}{2\eta} ~g_{+-}(\eta)]
G^{+}_{00} ~~,  \eqno (10a)
$$

$$
G_{M}  = \frac{1}{2p^{+}(2\eta + 1)} [2 + \frac{2\eta - 1}{\eta} ~
g_{+0} (\eta) - \frac{1}{\eta}~ g_{+-}(\eta)] 
G^{+}_{00} ~~
{\hskip 2.5cm} \eqno (10b)
$$
and
$$
G_{Q}  = \frac{1}{2p^{+}(2\eta + 1)} [-1 + \frac{1}{\eta} ~
g_{+0}(\eta) - \frac{\eta + 1}{2 \eta^2} ~g_{+-}(\eta)]
G^{+}_{00} ~~. {\hskip 2.5cm}  \eqno (10c)
$$

Now we discuss the constriants on $g_{+0}(\eta)$ and $g_{+-}(\eta)$
from the experimental data. As we know, $G_{M}(Q^{2})$ changes sign at
$Q^{2} = Q^{2}_{0} \sim 2 GeV^{2~[10]}$ (or $\eta = \eta_{0} = \frac{Q^{2}_{0}}
{4M^{2}} \simeq 0.13)$. The dominance of $G^{+}_{00}$ can not explain this
point (see Eq.(3b)) and at least the second term in Eq.(3b) should be in the
same order of the first term to cancel it in order to fit data of $G_{M}$.
That means $g_{+0}(\eta)$ is nonzero. If we first keep $g_{+-}(\eta)=0$ in Eqs.(10),
then $g_{+0}(\eta_{0})$ can be determined by the zero at $G_{M}(\eta_{0})$,
which turns out to be $g_{+0}(\eta_{0}) = \frac{2\eta_{0}}{1-2\eta_{0}}$.
In this case, $G_{Q}$ is negative at $\eta = \eta_{0}$ where PQCD begins to be
valid. Thus there must be a node in the region $Q^{2} < 1 GeV^{2}$ since $G_{Q}$
is positive at the origin experimentally. The theoretical prediction is contrary
to the experimental data without $G^{+}_{+-}$ contribution. Therefore the predicted
form factors are in sharp disagreement with the data without $G^{+}_{+0}$ and
$G^{+}_{+-}$ contributions and the existence of non-zero $g_{+0}(\eta)$ and
$g_{+-}(\eta)$ is necessary. In addition to the constrant (9), $g_{+0}(\eta)$
and $g_{+-}(\eta)$ should astisfy

$$
g_{+0}(\eta_{0}) = \frac{2\eta_{0} - g_{+-}(\eta_{0})}{1-2\eta_{0}} \eqno (11)
$$
to ensure $G_{M}(\eta_{0}) = 0$ and

$$
2\eta ~g_{+0}(\eta) - 2\eta^{2} < (\eta + 1) ~~g_{+-}(\eta)  \eqno (12)
$$
to keep $G_{Q}$ positive at any momentum transfer. In particular, as $\eta
= \eta_{0}$ Eqs.(11) and (12) give the constraints on $g_{+0}(\eta_{0})$ and
$g_{+-}(\eta_{0})$,
$$
g_{+-}(\eta_{0}) > \frac{2\eta^{2}_{0}}{1-\eta_{0}} \eqno (13)
$$
and
$$
g_{+0}(\eta_{0}) < \frac{2\eta_{0}}{1-\eta_{0}} \eqno (14)
$$
Eqs.(9), (11) and (12) are three constraints on the functions $g_{+0}(\eta)$
and $g_{+-}(\eta)$.

\section{A Phenomenological Example}

{\hskip 0.6cm}
As mentioned in the section 2, $g_{+-}(\eta) \neq 0$ is important to keep
$G_{Q} > 0$ although $G^{+}_{+-} = \frac{1}{2\eta} g_{+-}(\eta) G^{+}_{00}$ is suppressed
by the higher order factor $\Lambda^{2}_{QCD}/Q^{2} ( = \frac{\Lambda^{2}_{QCD}}
{2M^{2}} \cdot \frac{1}{2\eta})$. However the $G^{+}_{+0}$ is the first order
correction which makes the zero in $G_{M}$ at $Q^{2}_{0} \simeq 1.85 GeV^{2}$.
We have expanded $G^{+}_{+0}$ and $G^{+}_{+-}$ to the second order in
$\Lambda_{QCD}/M$ in Ref.[6] and numerical results show that the second order
contribution to $G^{+}_{+0}$ plays an important role in the
intermediate energy region. In this paper we introduce an exponential form phenomenologically
as an example,
$$
g_{+0}(\eta) = f ~~exp(-\frac{bf}{\sqrt{2\eta}} ) \eqno (15a)
$$
and
$$
g_{+-}(\eta) = f^{2} exp(-\frac{cf}{\sqrt{2\eta}} ) \eqno (15b)
$$
to interpolate the higher order corrections. Obviously, the exponential
form (15) satisfies Eq.(9) and is consistent with the perturbative QCD
prediction at the large transfer momentum region. Thus Eq.(15) is enable  us
to make an analytical evaluation to the deuteron form factors.

\section{Numerical Results}

{\hskip 0.6cm}We input the parameters $b$ and $c$, and determine $f$ by the zero in $G_M$.
For a certain $c$, the obtained $G_Q$ increases proportionally with $b$.
We can fix $b$ to connect our predictions with the data smoothly.
To retain good convergence, we constrain 
$b\frac{f}{\sqrt{2\eta}}$ and $c\frac{f}{\sqrt{2\eta}}$ being smaller
than unity. 
On the other hand, we restrict $b$ and $c$ to be positive to keep the 
exponential damping as $Q^2$ goes to infinity and it is a resonable assumption,
 after taking into account
the higher order corrections.
For $c=0.0,0.5$ and -0.5,
the predicted $B(Q^2)$, $G_Q(Q^2)$ and
$T_{20}(Q^2)$ are shown in figs.~(1-3). 
Experimental data are taken from
Refs.[1,10-12].
To smoothly connect with the data of $G_Q$, the parameter $b$ should be
1.1, 1.3, and 0.8, $f=0.37, 0.51$, and 0.30, respectively. 
As argued above, $c=-0.5$ should be abandaned.
While $c=0.5$, the corresponding $f$ is 0.51, which is too large to keep
$b\frac{f}{\sqrt{2\eta}}$ smaller than unity as $Q^2\geq 1$.
The parameter $c=0.0, b=1.1, f=0.37$ is an appropriate choice.
 
\section{Summary}

{\hskip 0.6cm}Based on perturbative QCD predictions at large momentum transfers we have
tried to discuss the corrections to deuteron form factors beyond the second
order in $\Lambda_{QCD}/M$.A general consideration is given by introducing
 functions $g_{+0}(\eta)$ and  $g_{+-}(\eta)$ and the data at the present energy
 region put three constrants on the functions  $g_{+0}(\eta)$ and  $g_{+-}(\eta)$.
 In order to explore the role of higher order
contributions we suggest an exponential form for $g_{+0}(\eta)$ and $g_{+-}(\eta)$ 
as an example. We conclude that (1) the helicity-zero to
zero matrix element $G^{+}_{00}$ dominates the gross structure function $A(Q^2)$
in both of the large and intermediate energy regions. A QCD-inspired model
can describe this matrix element well$^{[3]}$. (2) $G^{+}_{+0}$ and $G^{+}_{+-}$
contributions are
important in determining form factors $G_M$ and $G_Q$; 
(3) By fitting the data, we get a set of parameters
in $g_{+0}(\eta)$ and $g_{+-}(\eta)$, 
$c=0.0, b=1.1, f=0.37$, which can describe
$G_M,G_Q$ and $T_{20}$ appropriately.
the parameters $c=0.0$ indicates  that it is good for $G^{+}_{+-}$
in the intermediate energy region to take the asymptotic behavior which were
predicted by pQCD, and the higher order (beyond the second order)
contributions are negligible; (4) The higher order corrections to $G^{+}_{+0}$
should be taken into account and they make sizeable contributions in the
intermediate energy region.


\newpage
\section*{Figure Captions}
\begin{description}
\item[Fig.1.] Structure function $B(Q^2)$. 
The dashed dotted line corresponds to the Paris potential calculation.
Experimental data are taken from Ref.[10].
\item[Fig.2.] The form factor $G_{Q}$.
Experimental data are taken from Ref.[11].
\item[Fig.3] The tensor polarization $T_{20}$ with scattering angle $\theta = 70^\circ$.
The dashed dotted line corresponds to the
calculation with Paris potential. 
Experimental data are taken from Ref.[11]
\end{description}
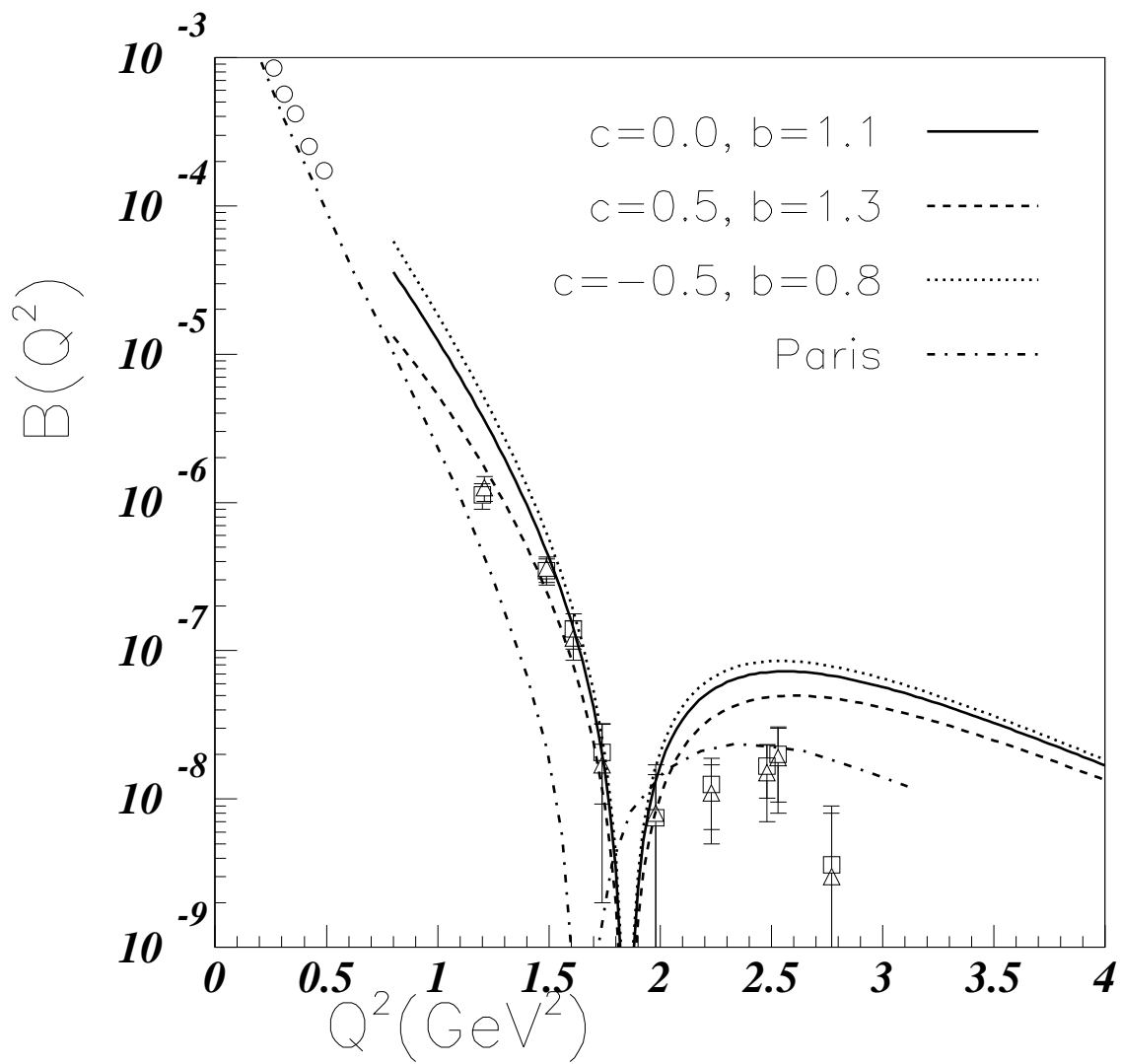
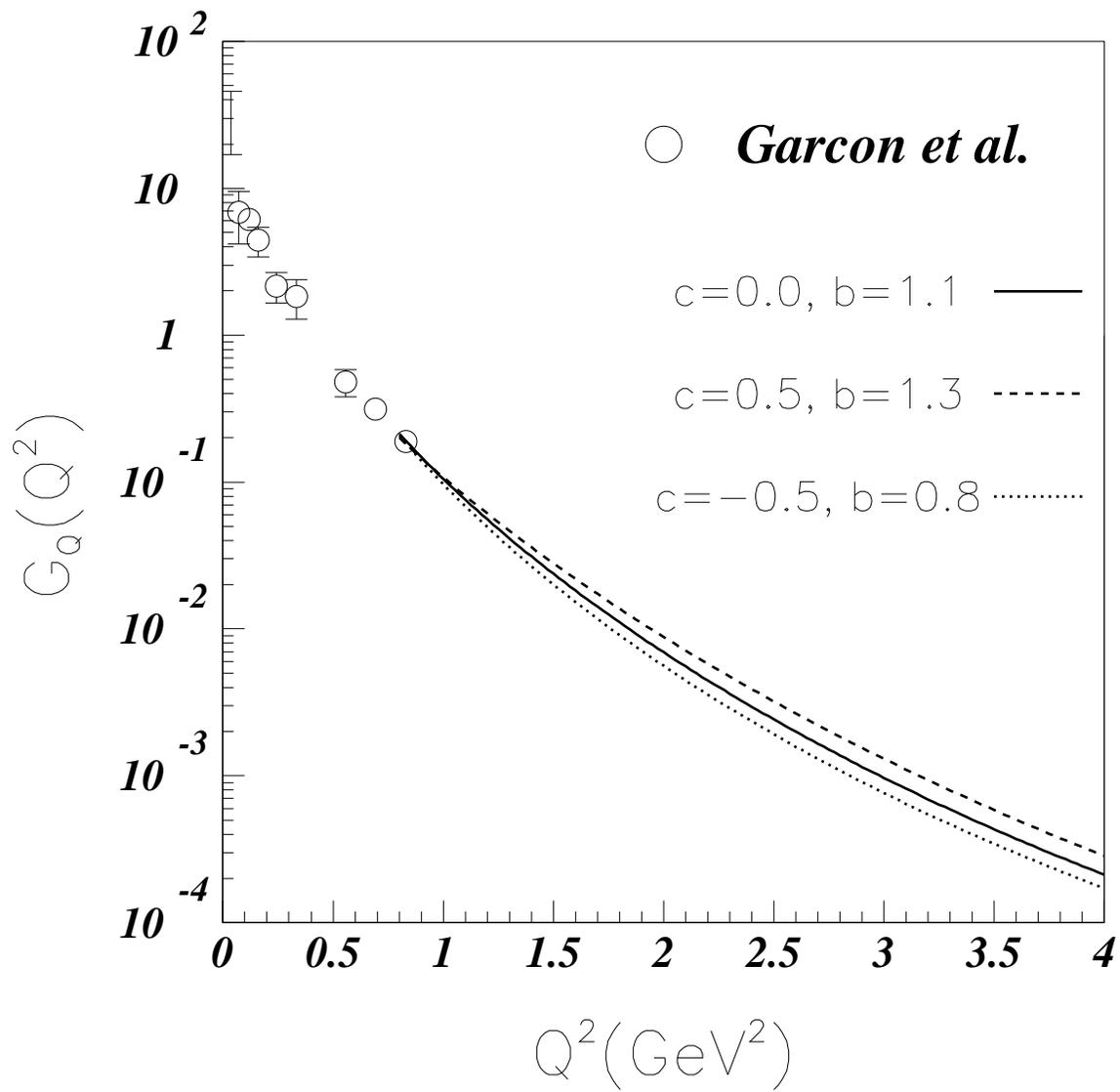
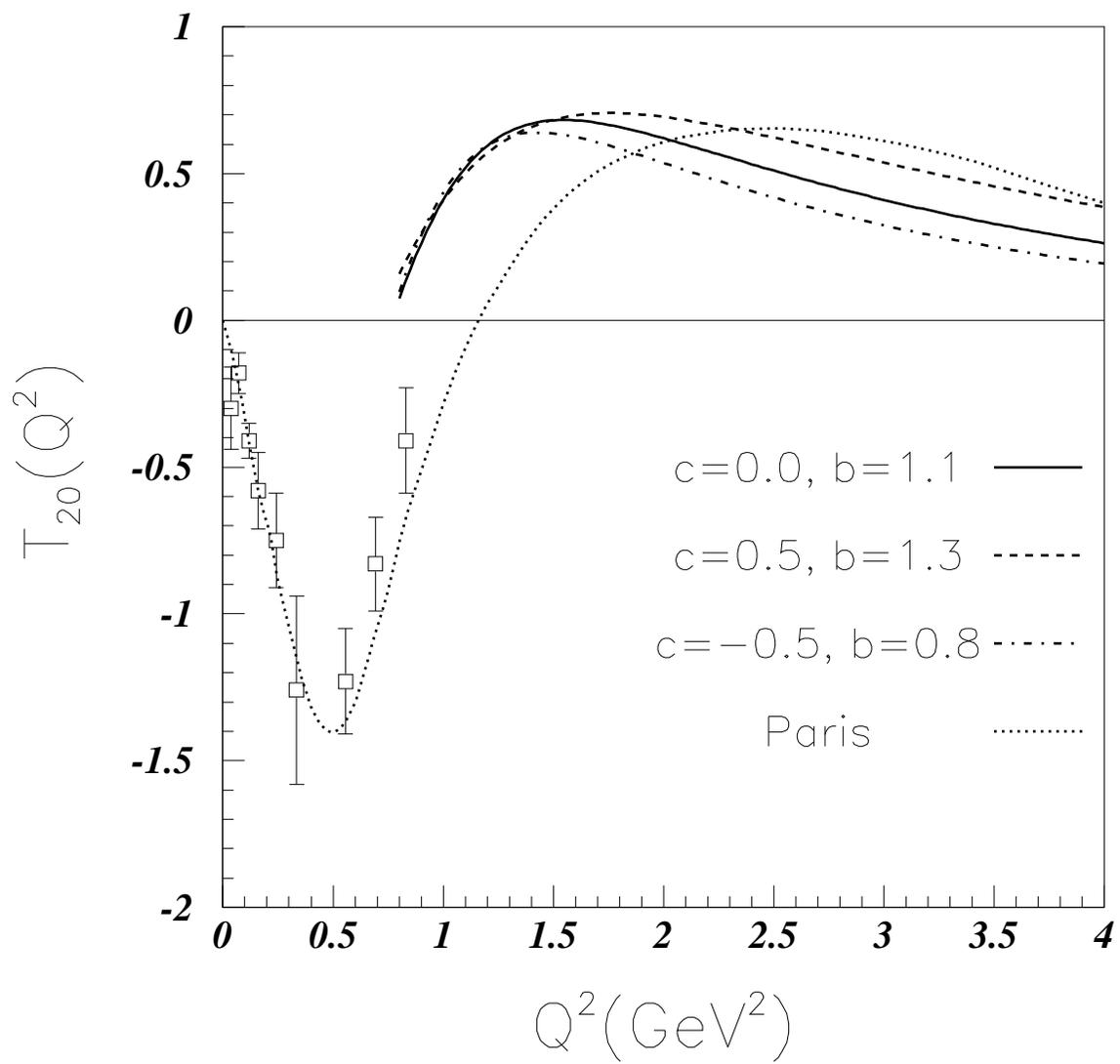
\end{document}